\documentclass[prd, nofootinbib]{revtex4-1}
\usepackage{amssymb}

\usepackage{amsmath}

\usepackage{graphicx}

\usepackage{amsfonts}

\newcommand{\be}{\begin{equation}}
\newcommand{\ee}{\end{equation}}
\newcommand{\beq}{\begin{eqnarray}}
\newcommand{\eeq}{\end{eqnarray}}
\newcommand{\bea}{\begin{eqnarray}}
\newcommand{\eea}{\end{eqnarray}}
\def\a{\alpha}

\begin{document}

\title{%
Noisy soccer balls}
\author{Giovanni Amelino-Camelia$^a$, Laurent Freidel$^c$, Jerzy Kowalski-Glikman$^b$, Lee Smolin$^c$
\\{\it
$^a$Dipartimento di Fisica, Universit\`a ``La Sapienza" and Sez.~Roma1 INFN, P.le A. Moro 2, 00185 Roma, Italy\\
$^b$Institute for Theoretical Physics, University of Wroclaw,  Pl. Maxa Borna 9, 50-204 Wroclaw, Poland\\
$^c$Perimeter Institute for Theoretical Physics, \\ 31 Caroline
Street North, Waterloo, Ontario N2J 2Y5, Canada }}
\date{\today}

\begin{abstract}
In her Comment \cite{sabicomment} Hossenfelder proposes a
generalization of the results we reported in Ref.\cite{soccerREF}
and argues that thermal fluctuations introduce incurable pathologies
for the description of macroscopic bodies in the relative-locality
framework. We here show that Hossenfelder's analysis, while raising a very interesting point,
 is incomplete  and leads to incorrect conclusions.
Her estimate for the fluctuations did not take into account
some contributions from the geometry of momentum
space which must be included at the relevant order of approximation. Using the full
expression here derived one finds that thermal fluctuations are not in
general large for macroscopic bodies in the relative-locality
framework. We find that such corrections can be unexpectedly large
only for some choices of momentum-space geometry, and we comment on
the possibility of developing a phenomenology suitable for possibly
ruling out such geometries of momentum space.
\end{abstract}
\maketitle

\section{Introduction}
In a recent paper \cite{soccerREF} we considered the description
of macroscopic bodies within the relative-locality framework based on
curved-momentum-space.  Our main result was to show that in
an idealized case, in which the constituents of a macroscopic body are modelled as being in rigid motion, an apparently
troublesome aspect of the description of macroscopic bodies, the
so-called ``soccer-ball problem", is actually absent. This is
essentially because the deformation scale reflecting curvature of
momentum space, which is assumed to be of order $M_P$ for a single constituent, scales like
$N M_P$ for a body with $N$ constituents in rigid motion.   This was
reassuring, but for full assurance we should consider the case in
which the momenta of the constituents fluctuate around their mean.
This can be called the quasi-rigid motion.
 In the closing remarks of Ref.\ \cite{soccerREF}
we offered a simple argument encouraging the hypothesis that also
for bodies in quasi-rigid motion our solution of the soccer-ball
problem should stand. However, in a Comment \cite{sabicomment},
Hossenfelder criticized our argument.  Using  a
random-walk model of thermal fluctuations in a macroscopic body,
 Hossenfelder claimed that non-linear terms arising from the curvature of momentum space can accumulate leading to the apparently paradoxical
 result that the fluctuations overwhelm the contributions of averages.
If this was true the case of constituents in quasi-rigid motion
would be unmanageably pathological.

 Here we address this significant challenge to the cogency of the relative locality hypothesis by giving a more precise analysis of
the relevant thermal fluctuations.

We find, first of all,  that the question posed by Hossenfelder is
interesting and leads to a new type of investigation in which
thermal fluctuations probe the geometry of momentum space away from
 small momenta.  This may potentially open up new opportunities
to constrain the geometry of momentum space from the properties of
macroscopic bodies.

Our analysis leads to the identification of  three
classes of momentum space geometries.  There are cases in which
Hossenfeldler's worry is realized and the fluctuations of the
momenta of the constituents of a hot macroscopic body  have effects
which scale like positive powers of $N$ in ways that seem to
contradict our observations.  On the other hand there are also
many cases in which the dangerous terms cancel and the
fluctuations of the deformed theory behave like those of standard
statistical physics. Most interesting is, however, the
third, in between, case, in which there may be new physical effects
by which the thermodynamics of large bodies may be able to serve as
a probe of the geometry of momentum space for macroscopic values of
momentum and energy.

Our analysis is more careful than both our brief remarks in
\cite{soccerREF} and Hossenfelder's approach \cite{sabicomment},
which did not keep all terms necessary to do a consistent
perturbative analysis of the effects of the fluctuations.  Taking
all terms into account  at each order means that we can constrain
the geometry of the addition rule
 to higher order that was possible before, and still leave us with a open set of non trivial possibilities.

\section{Total-momentum fluctuations within a consistent perturbative approach}
We take as starting point $P^{(N)}\equiv
\oplus_{a=1}^{N}(\bar{p}+\delta p^{a}) $ to be  the total momentum
of our soccer ball in the relative-locality framework. Here,
$\bar{p}$ is the average value of the momenta for each constituent,
and $\delta p^{a}$ is the fluctuation of this value for the
constituent $a$. The precise definition is by recurrence $P^{(N+1)}=
P^{(N)}\oplus (\bar{p} +\delta p^{N+1}),$ with $P^{(0)}=0$. The
deformed addition rule $\oplus$ is a manifestation of the geometry
of momentum space, which, as done in \cite{sabicomment}, we describe
in terms of a perturbative series in inverse powers of $\/M_P$: \be
(p\oplus q)_{\a}= p_{\a} + q_{\a} + M_{P}^{-1}\Gamma_{\a}{}^{\mu\nu}
p_{\mu}q_{\nu} + M_{P}^{-2}\Delta_{\a}{}^{\mu\nu\rho}
p_{\mu}p_{\nu}q_{\rho} +
M_{P}^{-2}\tilde{\Delta}_{\a}{}^{\mu\nu\rho} q_{\mu}q_{\nu}
p_{\rho}+ \cdots \label{quadratic} \ee Notice that we included terms
up to quadratic order in $\/M_P$, whereas Hossenfelder only
considered the linear order. We shall soon see that a correct
estimation of the size of fluctuations of the total momentum of a
body reveals that the leading contribution is quadratic in $\/M_P$,
so that truncations of the composition law to linear order, such as
Hossenfelder's, cannot give a consistent description.

Also note that we did not include in (\ref{quadratic}) terms of the form $\Gamma pp$ or $\Delta ppp$ or $\Delta qqq$ because,
 as already done in \cite{soccerREF}, \cite{sabicomment}, we  work in connection-normal coordinates.
This are the coordinates in which the composition laws is linear for collinear momenta, $(a p \oplus b p)_{\mu} = (a+b) p_{\mu}$
(for any real number $a,b $). It also
implies that the following conditions must be satisfied:
\be\label{Normalcoord}
\Gamma_{\a}^{(\mu\nu)} =0,\qquad \Delta_{\a}{}^{(\mu\nu\rho)} +\tilde{\Delta}_{\a}{}^{(\mu\nu\rho)}=0,
\ee
where the bracket means symmetrization of indices.

For the analysis of how small fluctuations of the momenta of the constituents
affect the total momentum of the body we consider contributions
to the total momentum which are linear in the fluctuations:
\be
P_{\alpha}^{(N)}= N\bar{p}_{\alpha} + \sum_{a} (W_{a}^{(N)}(\bar{p}))_{\alpha}^{\beta} \delta p^{a}_{\beta} + \cdots,
\label{forW}
\ee
where the dots stands for higher-order terms in $\delta p$. The first term takes the simple form $N\bar{p}_{\alpha}$
thanks to the adoption of connection-normal coordinates.
$W_{a}^{(N)}(\bar{p})$ in the second term is the coefficient of $\delta p^{a}$ in the expansion of  $P_{\alpha}^{(N)}$
and can be explicitly calculated from the knowledge of $\oplus$.
In particular on the basis of (\ref{quadratic}) our task will be to compute $W_{a}^{(N)}$ to quadratic order
in $1/M_P$.

For genuine fluctuations we  demand that  the second term in
(\ref{forW}) is much smaller than the first one and  average to zero in the large $N$ limit.
 Hossenfelder's observes that it is not enough to demand that the
 fluctuations are small. For consistency we need to evaluate the size of fluctuations
around the mean value and show that it is negligible as well.

The size of fluctuations depends crucially on  $W_{a}^{(N)}$ through
\be
(\sigma^2)_{\alpha \beta}
= \frac{1}{N^{2}}\sum_{a,b}
\langle (W_{a}\delta p^{a})_{\alpha} (W_{b}\delta p^{b})_{\beta} \rangle
\ee where $\langle \cdots\rangle$ denotes the average that will be
described in the next section.

We definitely agree with Hossenfelder
that a fully satisfactory analysis of macroscopic bodies must
consider the size of fluctuations around the mean value. However, we
here show that the type of contributions to $\sigma^2$ on which
Hossenfelder's analysis focused, which have quadratic dependence on
the Planck scale, require an analysis of the composition law to the
 $M_{P}^{-2}$ order. As we shall show, the fact that
Ref.\ \cite{sabicomment} only included order-$M_{P}^{-1}$
corrections to the composition law led Hossenfelder to incorrect
results.

\section{Size of fluctuations in the non-relativistic regime}
Having characterized the needed properties of $W_{a}(\bar{p})$
we are now ready for estimating the size of the fluctuations.
For this we shall assume\footnote{We agree with Ref.\cite{sabicomment}
that the idealization of completely random fluctuations might be a good starting point
for the investigation of the problem at hand. Still it should be appreciated
that for some macroscopic bodies of definite interest this idealization is unrealistic.
In particular in a solid there are long-range binding forces  and a better approximation is to
assume gaussian fluctuations where $\langle \delta p^{a}_{i}\rangle =0$, but now
$\langle \delta p^{a}_{i}\delta p^{b}_{j}\rangle = \delta^{ab}\Delta_{ij}$, where $\Delta_{ij}$ is the
inverse interaction kernel. This kernel depends on where the atoms are in the solid.},
as done in Hossenfelder's Comment \cite{sabicomment}, that
 the momenta $\vec{p}$ fluctuate randomly,  $\langle \delta p^{a}_{i}\rangle =0$,
$\langle \delta p^{a}_{i}\delta p^{b}_{j}\rangle \propto \delta^{ab}\delta_{ij}$.

The point to be stressed is that the fluctuations must be
`physical", and when binding forces are neglected (as here and in
\cite{sabicomment})  the on-shellness of each constituent particle
must be preserved. This condition was not imposed, in
\cite{sabicomment} so in general the fluctuations described in
\cite{sabicomment} would be unphysical virtual fluctuations. We
shall here improve the analysis also in this respect.

Since the bulk of our experimental knowledge of the properties of
macroscopic bodies is in the non-relativistic regime, we shall focus
on the case where $\bar{p}_{j} \ll \bar{p}_{0} \simeq m$ (and assume
again for simplicity the all constituents have the same mass $m$).
Here we assume for simplicity that in the non-relativistic limit
the on-shellness relation  for constituent particles is undeformed
to second order.
 This means that we can take
$e\equiv\bar{p}_{0} = m +\vec{p}^{2}/(2m)$ and $e\delta e -
\vec{p}\delta \vec{p}=0$, i.e. $ \delta e = \bar p_i\delta p_i/e.$
In the non-relativistic limit we can rely  on the Boltzmannian
distribution, with the one particle probability density given by $
f(p) \sim \exp(- \frac{p^{2}}{2mT})$, and therefore \be\label{fluct}
 \langle \delta p_{i}^{a}\delta p_{j}^{b}\rangle = m T \delta^{ab} \delta_{ij}.
\ee The fluctuation in energy follows from the mass-shell relation
$\delta e = u_{i} \delta p_{i}$ where $u_{i}=p_i/m$ is the spatial
velocity. Even if we are in the non relativistic limit we can use a
relativistic notation to write down the fluctuation of both energy
and momenta,  we introduce a 4-velocity $u_{\a}= (1,u_{i})$ and we
can write the non relativistic  fluctuation as: \be \langle \delta
p_{\alpha}^{a}\rangle =0,\qquad \langle \delta p_{\alpha}^{a}\delta
p_{\beta}^{b}\rangle = mT\delta^{ab}\left(\eta_{\alpha \beta} +
2u_{\a}u_{\beta}\right). \ee
 We can now use this fluctuation model to evaluate the size of fluctuation to be given by the average
 $ N^{2} (\sigma^{2})_{\a\beta} = \langle \sum_{a,b}W_{a}^{(N)}(\bar{p})_{\alpha}^{\beta}  W_{b}^{(N)}(\bar{p})_{\alpha'}^{\beta'} \delta p^{a}_{\beta} \delta p^{b}_{\beta'}\rangle$. With the Boltzmannian model this gives
\bea
 (\sigma^{2})_{\a\a'} =\frac{m T}{N^{2}} \left< \left( \sum_{a} [(W_{a}^{(N)}(\bar{p}))^{T}\eta  W_{a}^{(N)}(\bar{p})]_{\alpha\alpha'}+ \frac2{m^{2}}
\sum_{a}
[W_{a}^{(N)}(\bar{p})\bar{p}]_{\alpha}[W_{a}^{(N)}(\bar{p})\bar{p}]_{\alpha'}\right)
\right> \label{joc} \eea where ${}^{T}$ denotes the transposition.
The first key point is that in our connection-normal coordinates the
second term in (\ref{joc}) vanishes identically: \be
[W_{a}^{(N)}(\bar{p})\bar{p}]_{\alpha} =0. \ee The size of
fluctuations is therefore simply given by \be (\sigma^{2}
)_{\a\a'}=\frac{m T}{N^{2}}  \left( \sum_{a}
[(W_{a}^{(N)}(\bar{p}))^{T}\eta
W_{a}^{(N)}(\bar{p})]_{\alpha\alpha'}\right)  \, . \ee The
goal is now to evaluate this object up to quadratic order.

\section{Pertubative expansion}
Let us first presents the results we obtain for the expansion of
$W_{a}(\bar{p})$ and the fluctuations to second order. Some relevant
results are collected in the Appendix. From the expansion
(\ref{19a}), the definition of $W_{a}^{(N)}(\bar{p})$, and
restricting our focus to a rest frame analysis, so that $\bar
p_{\a}= m\delta_{\a}^{0}$, we get \bea (W_{a}^{(N)})^\beta_\alpha=
\delta^\beta_\alpha - \frac{m}{M_{P}}(N+1-2a)
\Gamma_{\a}{}^{0\beta} + \frac{m^{2}}{M_{P}^{2}}\left( A_{a}^{N}
\Gamma^{0}\Gamma^{0} + B_{a}^{N} S^{00} + C_{a}^{N}
{\tilde{\Delta}}^{00}+D^{N}_{a} {\Delta}^{ 00 }\right)^\beta_\alpha.
\nonumber \eea where $A_{a}^{N},B_{a}^{N},C_{a}^{N},D^{N}_{a}$ are
$N$ dependent coefficients, explicitly derived in the Appendix.
Here we use the matrix notation
 $(\Gamma^{0})_{\a}^{\beta}\equiv \Gamma_{\a}{}^{0\beta}$, similarly $ (S^{00})_{\a}^{\beta}\equiv S_{\a}{}^{00\beta}$ etc...
 and we have defined $S_{\a}{}^{\beta \gamma \delta} \equiv 3 \Delta_{\a}{}^{(\beta \gamma \delta)}$.
Since $N$ can be very large for a macroscopic body it is
particularly important to keep track of the  order N of each term.
We find that the dominant contributions come from \be \sum_{a=1}^{N}
(N+1-2a)^{2}\approx \frac{N^{3}}{3},\qquad \sum_{a=1}^{N} B_{a}^{N}
\approx \frac{N^{3}}{3},\qquad \sum_{a=1}^{N} D^{N}_{a} \approx
\frac34 N^{2}. \ee We see that the $D$ term is only of order $N^{2}$
instead of $N^{3}$, The other sums involved in the evaluation of the
fluctuation are all of order less or equal to  $N^{2}$.

Equipped with this, we can now evaluate the the first term in
(\ref{joc}) finding that \bea\label{fluctuation}
\frac{(\sigma^2)_{\a  \a'}}{m^2} = \frac{ T}{N m}
 \delta_{\a  \a'}  - \frac{T }{M_{P}}  [(\Gamma^{0}\eta)^{T} + \eta\Gamma^{0}]_{\a  \a'}
 +  \frac{m T N}{3 M_{P}^{2}}\left[ (\Gamma^{0}\eta \Gamma^{0}) +[(S^{00}\eta)^{T}+ \eta S^{00}] +O(1/N) \right]_{\a  \a'}
\eea
 ${}^{T}$ denotes the transposition.

Let us pause briefly for comparing our constructive result to the
estimate proposed by Hossenfelder in \cite{sabicomment}. Within our
 notation that estimate would take the form
$$\frac{\sigma^2{}_{Hoss} }{m^{2}}\sim (\Gamma^{0})^{2} \frac{T^{2} N}{M_P^{2}}.$$
It should be noticed first of all that Hossenfelder's heuristic estimate does not
reproduce exactly any of the terms in the result (\ref{fluctuation}) we here derived.
The single term in Hossenfelder's estimate does agree roughly with our term $(\Gamma^{0}\eta \Gamma^{0})$ in (\ref{fluctuation}),
however the size of fluctuations is controlled by a $mT$ dependence in our case while she uses $T^{2}$.
More importantly the single estimate is not a consistent approximation of our result since it misses the contribution
going like $(S^{00}\eta)^{T}+ \eta S^{00}$, which is of the same order. This is of course connected to our earlier remarks
about the consistency of a perturbative approach to this sort of derivation: our $(\Gamma^{0}\eta \Gamma^{0})$ term originates
from the $O(M_P^{-1})$ contribution to the composition law while our $(S^{00}\eta)^{T}+ \eta S^{00}$
term originates from the $O(M_P^{-2})$ contribution to the composition law, but they both appear at the same order of
approximation of $\sigma^2$.\\
The difference is not only quantitative but also very importantly qualitative:
whereas Hossenfelder was using her  evaluation as motivation for excluding all composition laws
at leading order, we find that, even if one did insist on excluding  the corresponding
type of corrections, there is only a constraint on the relationship between
leading-order form of the composition law, affecting $\Gamma^{0}\eta \Gamma^{0}$,
and next-to-leading-order form of the composition law, affecting $(S^{00}\eta)^{T}+ \eta S^{00}$.

Having derived the full ($1/M_P^2$-order) expression we can also
analyze the structure of (\ref{fluctuation}) from a wider
perspective. The first term is the usual fluctuation term which goes
away in the large $N$ limit. The second term is proportional to the
linear evaluation of the non-metricity tensor $N_{ij}\equiv \Gamma_{i}{}^{0}{}_{j} +
\Gamma_{j}{}^{0}{}_{i}$. It corresponds to a correction to the usual
fluctuations which is suppress by a  very small factor $ {T
}/{M_{P}}$. If we demand the non-metricity to vanish this implies
that the connection coefficient are entirely determined by the
torsion.

The only problematic term in this expansion is the last term
proportional to $((\Gamma^{0})^{T}\eta \Gamma^{0}) +
[(S^{00}\eta)^{T}+ \eta S^{00}] $, it is of order $NmT/M_{P}^{2}$.
The demand that the fluctuations do not scale with $N$, results in a
condition for one of the second order coefficients given by
$((\Gamma^{0})^{T}\eta \Gamma^{0}) + [(S^{00}\eta)^{T}+ \eta S^{00}]
=0$, in components this simply reads \be \Gamma_{i}{}^{0}{}_{k}
\Gamma_{j}{}^{0 k} +2 S_{(i j)}{}^{00} =0, \label{result} \ee which
tells us that the symmetric part of $S^{00}$ is fixed, while the
rest of the components of $\Delta^{00}, \tilde{\Delta}^{00}$ are
free. When this condition is satisfied we do still have additional
corrections proportional to $\frac{mT}{M_{P}^{2}} \Delta^{00}$, but
evidently none of these contributions is problematic.

\section{Outlook}
We had shown in \cite{soccerREF} that, contrary to earlier naive
arguments, there is no interpretational challenges nor any
phenomenological paradoxes in the idealized case of body composed of
constituents in rigid motion.

We here showed that even for bodies whose
constituents are in quasi-rigid
motion the relative locality framework does not in general encounter any pathologies, extending the analysis
up to quadratic order in the (inverse of) the Planck scale for effects linear in the fluctuations
of the momenta of the constituents.
Within our analysis the description of macroscopic bodies is completely unproblematic for the large class of non-trivial
momentum-space geometries that satisfy the condition (\ref{result}).

It is interesting to contemplate the possibility that geometries
that do not satisfy condition (\ref{result}) might be ruled out
experimentally exploiting the $N$ dependence of $\sigma^2$. But
here, we could also advocate a more phenomenological approach. We
start by noticing that corrections of order $NmT/M_{P}^{2}$ are
still extremely small (smaller than one part in $10^{22}$) for an
actual soccerball at soccer-playing temperatures of about 300
Kelvin, so we can confirm that the original   `soccer-ball problem'
\cite{soccerREF} is not going to be a matter of contention with any
choice of momentum-space geometry. As suggested by Hossenfelder
there might still be a problem for macroscopic bodies in some
extreme cases, such as ultra-hot and ultra-massive bodies studied in
astrophysics \cite{sabicomment}. But instead of jumping  to the
conclusion that mere existence of such astrophysical bodies suffices
for excluding some momentum-space geometries, we feel a more
prudently scientific approach would be to seek the opportunity to
analyze in future works an actual measurement result on such bodies
which is affected by large corrections, for suitable choices of
momentum-space geometry. A first step toward this goal would be to
generalize the analysis we here reported so that it could apply to
case with $T \gtrsim m$ (within the nonrelativistic limit, on which
we here focused, only cases with $m \ll T$ can be consistently
studied). Moreover, it will be necessary to gain some control also
about the interactions among macroscopic bodies in the
relative-locality framework: in principle for some momentum-space
geometries large corrections might be present for the abstract
notion of total momentum of a body and yet not be present in our
measurements of properties of that body, which inevitably require a
role for interactions with the body. This is clearly a line of
research worth pursuing since the opportunity of excluding at least
some geometries of momentum space would be extremely valuable for
the relative-locality research program.

Another interesting challenge we leave for future studies concerns the geometric interpretation
of results such as our condition (\ref{result}).
It is very tempting to assume that conditions such as (\ref{result}), when analyzed beyond the level of dry relationships
among tensors, might be characterizing in meaningful geometric way desirable (or needed) properties of momentum-space
geometry.
In closing we sketch out an argument suggesting that this could be the case and which opens new investigations.

We know from general theory of relative locality that $W_{a}^{(N)}$ is a transport operator from $0$ to $N\bar{p}$.
we also know that this transport operator is determined to first order by a connection evaluated at $0$.
Let us assume, for the sake of our argument, that this is true beyond first order.
That is lets suppose that $W_{a}^{(N)}$ can be written as the parallel transport of a connection $\Gamma$ along a path,
labelled by $a$, from $0$ to $N\bar{p}$.
That is we assume that $W_{a}(\bar{p}) =P_{a}\exp\int_{0}^{N\bar{p}} \Gamma $, where $P_{a}$ denotes the path ordering
along a path $a$. Even if that might not be generally true, assuming it gives us a key insight.

Let us also suppose  that the momentum space geometry is such that the
 connection is compatible with the
metric, that is that is $\nabla \eta =0$, in other words the
connection is the Levi-Civita connection plus  torsion possibly.
Integrating out this equation can be straightforwardly done and
leads to the following metric compatibility condition on the
transport operator: \be [(W_{a}^{(N)}(\bar{p}))^{T}\eta
W_{a}^{(N)}(\bar{p})]_{\alpha\alpha'} = \eta_{\a\a'}(N\bar{p}) \ee
where $\eta(p)$ is the metric at $p$ while $\eta= \eta(0)$. This
suggest that, under the conditions we specified,
 the fluctuations would be exploring the geometry of momentum space away from the origin.
This would lead to the exciting opportunity of exploring the geometry of momentum space away from the origin
by harnessing the power of the large number $N$.
It also opens a new set of mathematical investigation on when our hypothesis can be realize.

\appendix
\section{Properties of $W_{a}(\bar{p})$ to order $M_{P}^{-2}$}\label{details}
Here we  establish the dependence of $W_{a}(\bar{p})$ on the form of the composition law
to second order in (inverse of) the Planck scale.
It is useful to start by observing that from (\ref{quadratic}) we get:
\bea
[(a\bar{p}+\delta p)\oplus (b \bar{p} \oplus \delta q)]_{\a} =
(a+b) \bar{p}_{\a} + (\delta p + \delta q)_{\a} +
M_{P}^{-1} a\Gamma_{\a}{}^{\bar{p} \mu} \delta q_{\mu} + M_{P}^{-1} b\Gamma_{\a}{}^{ \mu \bar{p}} \delta p_{\mu}  \\
+ M_{P}^{-2} (ab\Delta_{\a}{}^{\mu \bar{p} \bar{p}}+ ab\Delta_{\a}{}^{\bar{p} \mu  \bar{p}} + b^{2} \tilde{\Delta}_{\a}{}^{ \bar{p} \bar{p} \mu} )\delta{p}_{\mu} \nonumber
+M_{P}^{-2} (ba\tilde{\Delta}_{\a}{}^{\mu \bar{p} \bar{p}}+ ba\tilde{\Delta}_{\a}{}^{\bar{p} \mu  \bar{p}} + a^{2} {\Delta}_{\a}{}^{ \bar{p} \bar{p} \mu} )\delta{q}_{\mu} + \cdots
\eea
where we used the notation $T^{\mu \bar{p} \nu}\equiv T^{\mu \a \nu}\bar{p}_{\a}$.
We can conveniently rewrite this expression by defining
$$S_{\a}{}^{\beta\mu\nu}= 3\Delta_{\a}{}^{(\beta\mu\nu)}.
$$
Using the fact that we are working with connection-normal coordinates, finding
\bea
[(a\bar{p}+\delta p)\oplus (b \bar{p} \oplus \delta q)]_{\a}  &=&(a+b) \bar{p}_{\a} + (\delta p + \delta q)_{\a} ~~~~~~~~
~~~~~~~~~~~~~~~~~~~~~~~~~~~~~~~~~~~~~~~~~~~~~~~~~~~~~~~~~~~~~~~~~~~~~~~~~~~~~~~~~~~~~~\nonumber\\
& & + \Gamma_{\a}{}^{\bar{p} \mu} (- b \delta p + a \delta q)_{\mu}
 + ab S_{\a}{}^{\bar{p}\bar{p}\mu}(\delta p - \delta q)_{\mu}  +
(b  {\tilde{\Delta}} -  a {\Delta})_{\a}{}^{ \bar{p} \bar{p} \mu}  (b \delta{p}- a\delta{q})_{\mu}.
\eea
From this we can now get the recursion equation determining the perturbation  to order $M_{P}^{-2}$:
\bea
[W_{a}^{(N+1)}]_{\a}^{\mu} &=& \left(\delta_{\beta}^{\mu} - \Gamma_{\a}{}^{\bar{p} \beta} +  N S_{\a}{}^{\bar{p}\bar{p}\beta}
+ (  {\tilde{\Delta}}_{\a}{}^{ \bar{p} \bar{p} \beta} -  N {\Delta}_{\a}{}^{ \bar{p} \bar{p} \beta})\right) [W_{a}^{(N)}]_{\beta}^{\mu},\\
{[}W_{N+1}^{(N+1)}]_{\a}^{\mu} &=& \delta_{\a}^{\mu} +N \Gamma_{\a}{}^{\bar{p} \mu} - N S_{\a}{}^{\bar{p}\bar{p}\mu}
-N(  {\tilde{\Delta}}_{\a}{}^{ \bar{p} \bar{p} \mu} -  N {\Delta}_{\a}{}^{ \bar{p} \bar{p} \mu}).
\eea
In the first line $ a=(1,\cdots,N)$, while the second line gives the initial condition for the recurrence.

This can be solved as \bea\label{19a} W_{a}^{(N)}= 1 - (N+1-2a)
\Gamma^{\bar{p}} +A_{a}^{N} \Gamma^{\bar{p}}\Gamma^{\bar{p}} +
B^{N}_{a} S^{\bar{p}\bar{p}} + C^{N}_{a} {\tilde{\Delta}}^{ \bar{p}
\bar{p} }+ D^{N}_{a} {\Delta}^{ \bar{p} \bar{p} } +\cdots \eea where
$\Gamma^{\bar{p}}$ denotes the matrix $
(\Gamma^{\bar{p}})_{\a}^{\beta}= \Gamma_{\a}{}^{\bar{p} \beta}$, and
similarly for the other symbols, while the coefficients are \bea
A_{a}^{N} &=& (-1)^{N-a}(a-1), \\
B_{a}^{N} &=& \frac{N(N-1)}{2} - \frac{a(a+1)}{2} +1, \\
C_{a}^{N} &=& N+1-2a, \\
D^{N}_{a} &=& - \frac{N(N-1)}{2} + \frac{(3a-2)(a-1)}{2}.
\eea

\section*{ACKNOWLEDGEMENTS}

  We thank S. Hossenfelder for conversations and correspondence.   The work of G. Amelino-Camelia
was supported in part by a grant  from The John Templeton Foundation. The work of J. Kowalski-Glikman
was supported in parts by the grants  182/N-QGG/2008/0,
2011/01/B/ST2/03354 and by funds provided by the National Science
Centre under the agreement DEC-2011/02/A/ST2/00294.
L. Smolin's research is supported by grants from FQXi, NSERC and the Templeton Foundation.
L. Freidel's research is supported by a grant from NSERC.
  Research at Perimeter Institute
for Theoretical Physics is supported in part by the Government of
Canada through NSERC and by the Province of Ontario through MRI.


\begin{thebibliography}{0}

\bibitem{sabicomment}
  S.~Hossenfelder,
  ``Comment on arXiv:1104.2019, `Relative locality and the soccer ball problem,' by Amelino-Camelia et al,''
  arXiv:1202.4066 [hep-th].

\bibitem{soccerREF}
  G.~Amelino-Camelia, L.~Freidel, J.~Kowalski-Glikman and L.~Smolin,
  ``Relative locality and the soccer ball problem,''
  Phys.\ Rev.\ D {\bf 84} (2011) 087702
  [arXiv:1104.2019 [hep-th]].

\bibitem{principle}
  G.~Amelino-Camelia, L.~Freidel, J.~Kowalski-Glikman and L.~Smolin,
  ``The principle of relative locality,''
  Phys.\ Rev.\ D {\bf 84} (2011) 084010
  [arXiv:1101.0931 [hep-th]].



\end{thebibliography}
\end{document}